# Влияние магнитного поля на зарождение и свойства 0-градусных доменных границ в одноосных пленках с неоднородным магнитоэлектрическим взаимодействием.

**Вахитов Р.М., Солонецкий Р.В., Ибрагимова А.Р.**

[1]Уфимский университет науки и технологий, 450076, г. Уфа, Россия

vakhitovrm@yahoo.com

**Аннотация**

В работе исследуется поведение 0 –градусных доменных границ, возникающих в одноосных магнитных плёнках с флексомагнитоэлектрическим эффектом, в магнитном поле. Показано, что при определённых ориентациях магнитного поля можно существенно усилить (или ослабить) степень проявления флексомагнитоэлектрического эффекта в изучаемых плёнках. Кроме того, варьированием величины и направлениям магнитного поля можно значительно понизить (вплоть до нуля) значение критического электрического поля зарождения данной неоднородности. Установлено также, что 0 –градусная доменная граница неелевского типа, в которой индуцируемые связанные заряды не создают результирующего поля (экранировка полей), при наложении магнитного поля, направленного перпендикулярно плоскости вращения магнитных моментов, возникает флексомагнитоэлектрический эффект, усиливающийся с возрастанием величины магнитного поля.

## ВВЕДЕНИЕ

Исследования магнитоэлектрических явлений в магнитных материалах, как правило, вызывают повышенный интерес, что связано как с новыми необычными свойствами, обнаруженными в них, так и с возможностью их применения в различных устройствах спинтроники с низким энергопотреблением [1-3]. Одним из таких эффектов, привлекший внимание многих исследователей, стало обнаруженное в пленках ферритов-гранатов при комнатных температурах явление смещения доменных границ (ДГ) под действием неоднородного электрического поля [4]. На основе анализа экспериментальных данных авторами было высказано предположение, что наблюдаемое явление имеет магнитоэлектрическую природу и обусловлено наличием в изучаемых пленках неоднородного магнитоэлектрического взаимодействия (флексомагнитоэлектрический эффект), впервые предложенного к рассмотрению в работе [5]. Однако такая интерпретация опытных данных была подвергнута сомнению в работе [6], в которой был предложен другой механизм для объяснения обнаруженного явления. В его основе лежит эффект возможного смещения однотипных ионов под действием электрического поля заряженной иглы, поднесенной к поверхности пленки, что может привести к локальному изменению константы одноосной анизотропии. Сравнительный анализ обоих механизмов показал, что на качественном уровне они объясняют практически все известные экспериментальные результаты [7]. Отсюда следует, что рассмотренные механизмы не противоречат друг другу и каждый вносит свой вклад в проявлении данного эффекта. Однако какой из них является доминирующим можно выявить лишь в ходе дальнейших исследований. Тем не менее в работе [8], в которой изучались локальные электрические поля, генерируемые в ДГ в пленках ферритов-гранатов, на основе методики флуоресцентной спектроскопии одиночных молекул была подтверждена флексомагнитоэлектрическая (ФМЭ) природа наблюдаемого эффекта. Кроме того, изучение влияния зеемановского взаимодействия на степень проявления ФМЭ эффекта в пленках ферритов-гранатов показало [7,9], что, если внешнее магнитное поле направить перпендикулярно плоскости вращения магнитных моментов 180°ДГ, то, изменяя величину и направление поля, можно регулировать величиной эффекта (точнее, значением интегральной поляризации [7,9,10]) и даже «переключать» характер взаимодействия с притяжения на отталкивание. Последнее свойство может служить дополнительным аргументом в пользу ФМЭ природы рассматриваемого явления [11], а также иметь практическое применение. Следует отметить, что экспериментальные исследования данной ситуации [7,9] хорошо согласуются с результатами её теоретического обоснования на основе феноменологической модели [10,11]. В то же время в работе [12], где впервые были проведены расчеты возможных магнитных неоднородностей в одноосном ферромагнетике с неоднородным магнитоэлектрическим взаимодействием, было показано, что в таких материалах могут существовать наряду со 180° ДГ блоховского типа, (которая при приложении электрического поля становится квазиблоховской) еще и 0° ДГ с квазиблоховской структурой (0° ДГ (I)). Кроме того, дальнейшие исследования выявили, что в случае, когда на исследуемый магнетик действует неоднородное электрическое поле, то в области локализации поля может образоваться еще один вид магнитной неоднородности – 0° ДГ неелевского типа (0° ДГ (II)) [13]. Необходимо отметить, что неоднородности с такой нетривиальной структурой не являются чем-то необычным, несоответствующим реальным объектам в иерархии микромагнитных структур. Они относятся к семейству нетопологических солитонов [14] и при определенных условиях они могут стать устойчивыми образованиями, например, зародиться на определенного вида дефектах материалов [15-17]. Кроме того, они играют важную роль в спин-переориентационных фазовых переходах I рода, в которых 0° ДГ являются зародышем новой фазы, как бы осуществляя переход магнетика из одного состояния в другое посредством флуктационного механизма [18,19]), а также в процессах их намагничивания и перемагничивания, где они ассоциируются с доменами обратной намагниченности [20]. Следует отметить, что в случае ферромагнитных пленок с ФМЭ эффектом магнитные неоднородности типа 0° ДГ экспериментально не были обнаружены. Это вполне объяснимо, так как согласно [12] 0° ДГ (I) могут зародиться в подобных пленках лишь при больших величинах электрического поля, существенно превышающих их характерные значения, при которых ФМЭ эффект может проявиться, т.е. наблюдаться в эксперименте [4]. Поэтому представляет интерес изучение влияния магнитного поля на условия зарождения, структуру и поляризацию 0° ДГ в рассматриваемых пленках (как внешнего фактора с потенциальными возможностями изменения этих свойств).

.

**2. Основные уравнения**

Рассмотрим одноосный ферромагнетик с ФМЭ взаимодействием в виде пленки конечной толщины (D). Систему координат возьмем таким образом, чтобы ось Oz ∥ **n** (**n**- нормаль к поверхности пленки) и совпадала с легкой осью одноосной анизотропии, а ось Oy определяла направление, вдоль которого магнетик неоднороден, т.е. **M** = **M**(y), где **M** - вектор намагниченности (Рис.1). Вектор **M** удобно выразить через единичный вектор намагниченности **m**: **M** = $M_s$**m**, где $M_s$- намагниченность насыщения, **m**=$(\sin\theta\cos\varphi, \sin\varphi, \cos\theta\cos\varphi)$.

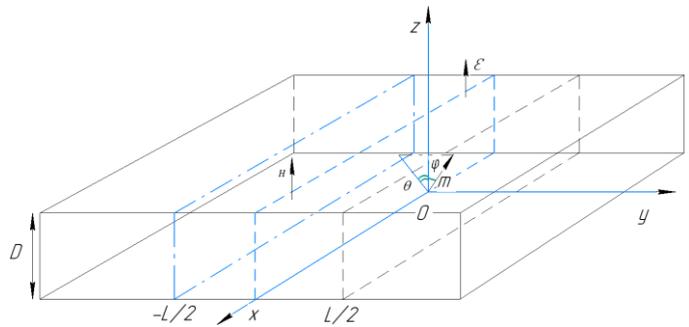

Рис.1. Схема, иллюстрирующая геометрию задачи

Тогда энергия такого магнетика, приведенная к площади сечения пленки плоскостью Oxz, будет иметь вид [11,12]

$$E = \int_{-\infty}^{\infty} \left\{ A\left[\left(\frac{d\varphi}{dy}\right)^2 + \cos^2\varphi\left(\frac{d\theta}{dy}\right)^2\right] + K_u(\sin^2\theta\cos^2\varphi + \sin^2\varphi) + 2\pi M_s^2 \sin^2\varphi + \varepsilon_{\text{н}} + \varepsilon_{\text{int}} \right\} dy \quad (1)$$

Здесь $\theta, \varphi$- полярный и азимутальный углы вектора **m** (Рис.1), A-обменный параметр, $K_u$- константа одноосной анизотропии, $\varepsilon_{\text{н}}, \varepsilon_{\text{int}}$- плотности энергии, соответственно, зеемановского и неоднородного магнитоэлектрического взаимодействий, которые можно представить в виде

$$\varepsilon_{\text{н}} = -M_s\mathbf{mH}, \quad \varepsilon_{\text{int}} = \mathcal{E}M_s^2\left[(b_1\cos^2\varphi + b_2\sin^2\varphi)\cos\theta\frac{d\varphi}{dy} + b_2\sin\theta\sin\varphi\cos\varphi\frac{d\theta}{dy}\right] \quad (2)$$

где $b_1, b_2$ – магнитоэлектрические постоянные [21], **H** и $\mathcal{E}$ – напряженности, соответственно, магнитного и электрического полей, при этом $\mathcal{E}\|Oz$, а **H** имеет произвольное направление. Кроме того, электрическое поле $\mathcal{E}$ считается неоднородным, т.е.

$$\mathcal{E} = \mathcal{E}_0/\text{ch}(y/L), \quad (3)$$

где L-ширина поперечного размера полосовой области действия электрического поля, $\mathcal{E}_0$- величина поля в центре полосы.

Предполагается, что рассматриваемая пленка является достаточно толстой ($\Delta_0 \ll D < \Lambda_0$, где $\Delta_0 = \sqrt{A/K_u}$ – характерная ширина 180° ДГ, $\Lambda_0 = \sqrt{A/2\pi M_s}$-размер линии Блоха [22]) и пренебрегается вкладом размагничивающих и рассеивающих полей.

Структура и свойства возможных магнитных неоднородностей описывается уравнениями Эйлера-Лагранжа, которые соответствуют минимуму (1) и имеют вид [4]:

$$\frac{d\varepsilon}{d\theta} - \frac{d}{dy}\frac{d\varepsilon}{d\theta'} = 0, \quad \frac{d\varepsilon}{d\varphi} - \frac{d}{dy}\frac{d\varepsilon}{d\varphi'} = 0, \quad (4)$$

где ε − представляет плотность полной энергии магнетика, которому отвечает подынтегральное выражение в (1), заключенному в фигурные скобки, штрихом обозначена производная по y.

Уравнения (4) представляют собой систему нелинейных дифференциальных уравнений с непостоянными коэффициентами, которую в данном случае можно решить только численными методами [11,12]. При этом для удобства будем пользоваться приведенными величинами: $\xi = y/\Delta_0$, $l = L/\Delta_0$, $Q = K_u/2\pi M_s^2$ (фактор качества), $h = M_s H/2K_u$, $\lambda = \mathcal{E}_0/\mathcal{E}_c$, $\mathcal{E}_c = 2K_u\Delta_0/M_s^2 b_0$ (характерная величина электрического поля), $b_0 = b_1 + b_2$, $\nu = p/p_0$, $p_0 = M_s^2 b_0/\Delta_0$ (характерная величина поляризации). Здесь $p = -d\varepsilon_{int}/d\mathcal{E}_0$ - дифференциальная поляризация.

### 3. 0° ДГ с квазиблоховской структурой

Исследование уравнений (2) при h=0 показывает, что решение, соответствующее 0° ДГ, возникает при достижении электрическим полем λ некоторого критического значения $\lambda = \lambda_n$ [12]. Зародившаяся таким образом 0° ДГ имеет неблоховское распределение намагниченности **m**, в котором угловая зависимость $\varphi = \varphi(\xi)$ представляет нечетную функцию (Рис.2,b). Последняя в момент зарождения в области $\xi > 0$ имеет один максимум и один минимум, и их значения ($\varphi_m$), взятые по абсолютной величине, не малы и уменьшаются при возрастании электрического поля λ. Сразу же при $\lambda > \lambda_n$ возникает еще один экстремум функции $\varphi = \varphi(\xi)$ при $\xi > 0$, соответствующий минимуму и возрастающий по абсолютной величине при увеличении λ [12].

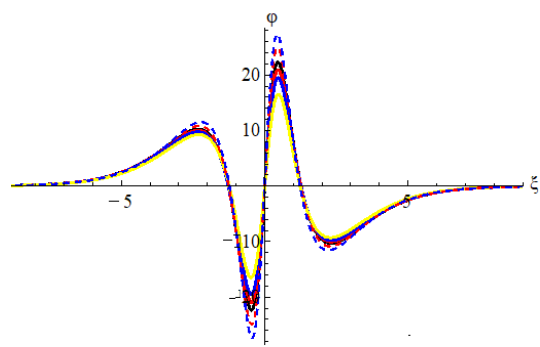

b)

Рис.2. Графики зависимостей углов θ (a) и φ (b) 0⁰ ДГ (I) от координаты ξ в магнитном поле H ∥ Ox, при λ = 4. Здесь и в дальнейшем берутся следующие значения материальных параметров : Q = 3, l = 5. Черная линия соответствует h = 0, красная - h = 0.2, синяя - h = 0.4, желтая h = 0.8, красная штриховая - h = -0.4, синяя штриховая - h = -0.8.

Очевидно, 0°ДГ (I) будет также иметь ненулевую поляризацию, весьма существенную в момент зарождения. Её интегральная величина в этот момент также значительна, но с увеличением поля она будет уменьшаться. Согласно расчетам, индуцированные в области ДГ связанные заряды распределяются по толщине стенки неравномерно, но симметрично относительно ее центра [12]. При этом в профиле распределения электрической поляризации имеет место три минимума, один из которых расположен в центре 0° ДГ ($\xi = 0$), и два максимума, расположенные на одинаковых расстояниях по обе стороны от центра ДГ (как и два других минимума). С возрастанием λ в области ДГ образуется тройной и даже пятерной электрический слой (при определенном значении λ [12]). В результате величина интегральной поляризации существенно уменьшается (Рис.3), т. к. создаваемые этими слоями электрические поля частично экранируют друг друга.

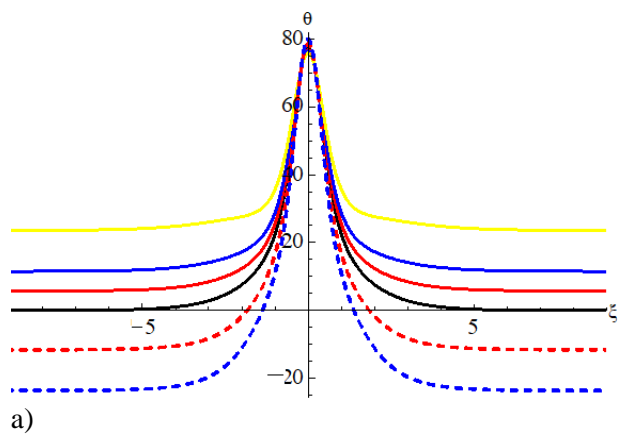

a)

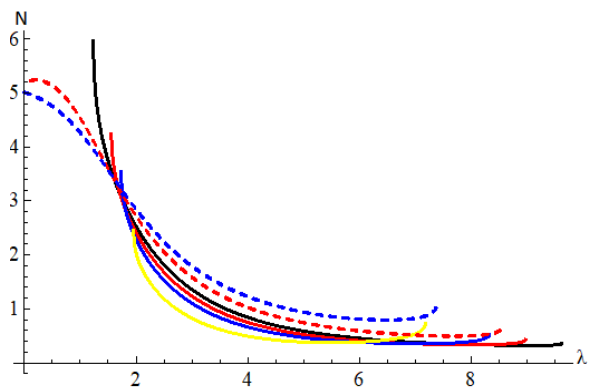

Рис.3. Графики зависимостей интегральной поляризации 0⁰ ДГ (I) от параметра λ в магнитном поле H ∥ Ox. Здесь черная линия соответствует h =

0, красная - h = 0.2, синяя - h = 0.4, желтая h = 0.8, красная штриховая - h = -0.4, синяя штриховая - h = -0.8.

Учтем теперь влияние внешнего магнитного поля на структуру и свойства $0°$ ДГ (I). (При включении магнитного поля с **H**∥Ox вектор намагниченности в доменах **M₀** отклоняется в направлении поля на угол $θ_0$, зависящий от величины h ($θ_0 = \arcsin(h)$) [11].Соответственно, максимальный угол отклонения $Δθ_m = θ_m − θ_0$, отчитываемый от однородного состояния, в зародившейся $0°$ ДГ (1) с возрастанием величины h будет уменьшаться. При этом максимальное значение $θ_m$ практически не меняется, в то же время $φ_m$ (по абсолютной величине) – уменьшается (Рис.2). Это приводит к тому, что экстремальные значения дифференциальной поляризации по абсолютной величине в каждом электрическом слое уменьшаются, что в свою очередь, приводит к понижению интегральной величины поляризации (Рис.3). Кроме того увеличивается и критическое поле $λ_n$ зарождения $0°$ ДГ (I). Таким образом, действие магнитного поля в рассматриваемой ситуации приводит к ослаблению ФМЭ эффекта.

В то же время, если направить поле **H** противоположно оси Ox, то изменение структуры $0°$ ДГ (I) будет таким же значительным по величине, но с противоположной тенденцией. В этом случае магнитные моменты, стремясь повернуться в сторону поля, приведут к наклону намагниченности **M₀** в доменах на угол $θ_0 = θ(∞) < 0$ и к возрастанию абсолютных величин экстремумов угловой зависимости $φ = φ(ξ)$ (Рис. 2). Это в свою очередь будет способствовать повышению абсолютных значений экстремумов дифференциальной поляризации $ν = ν(ξ)$. В то же время интегральная поляризация N=N(λ) ведет себя неоднозначно: в малых полях h (h< 1.6) она уменьшается при увеличении h, а в сильных полях (h> 1.6) – увеличивается (Рис. 3). Однако при этом имеет место другая особенность рассматриваемой ситуации: с возрастанием h резко уменьшается поле зарождения $0°$ДГ (I). В частности, при h= -0.8 будет отсутствовать порог зарождения $0°$ ДГ (I) по электрическому полю, что коррелирует с результатами исследований [19].

Если **H**∥Oz, то в этом случае при $λ ≠ 0$ симметрия распределения намагниченности в $0°$ДГ (I) не нарушается (Рис.4).

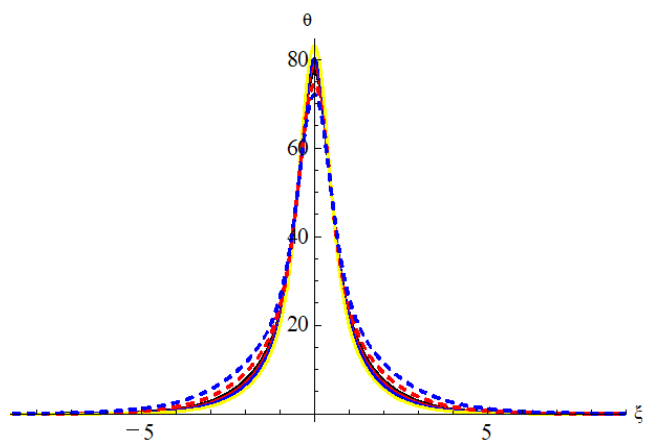

a)

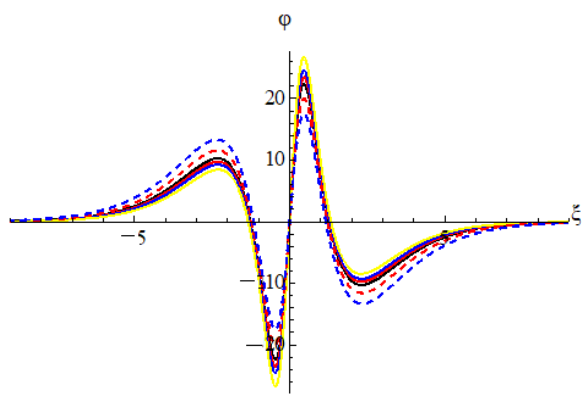

b)

Рис.4. Графики зависимостей углов θ (a) и φ (b) $0^0$ ДГ (I) от координаты ξ в магнитном поле H∥ Oz, при $λ = 4$. Здесь черная линия соответствует h = 0, красная - h = 0.2, синяя - h = 0.4, желтая h = 0.8, красная штриховая - h = -0.4, синяя штриховая - h = -0.8.

Зависимость $θ = θ(ξ)$ остаётся чётной функцией с одним максимумом в центре стенки, а $φ = φ(ξ) −$нечётной функцией с двумя минимумами и с двумя максимумами (Рис.4), которые при пространственной инверсии переходят друг в друга (если расположены на одинаковом расстоянии от центра ДГ). Однако при увеличении поля h увеличиваются максимальные значения углов $θ_m$ и $φ_m$ (при $ξ > 0$), в то время как минимальные углы $φ_m$ ($ξ > 0$) по абсолютной величине уменьшаются. Характер профиля распределения поляризации при этом также меняется: при малых λ представляет четверной электрический слой (из них средние слои одинаково заряжены, а крайние-противоположно), а при λ>2,96 (для h=0.4, Q=3) пятерной электрический слой (Рис.5).

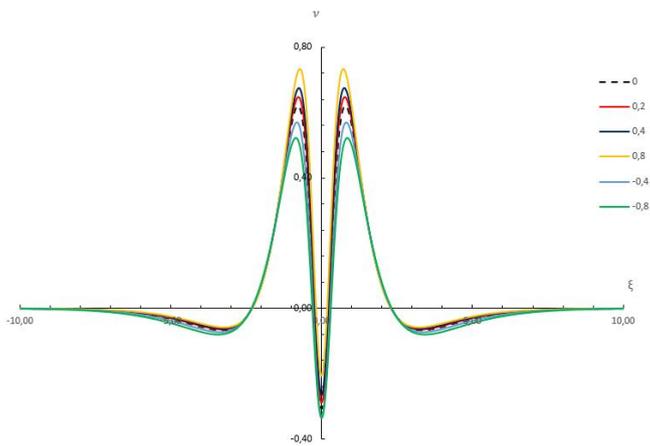

Рис.5. Графики, иллюстрирующие распределение дифференциальной поляризации ν в области 0° ДГ (I) для случая **H**∥Oz, при λ= 4. Здесь черная (штриховая) линия соответствует h=0, красная-h= 0.2, синяя-h = 0.4, оранжевая-h = =0.8, голубая-h = -0.4, зеленая-h = -0.8.

При возрастании h максимумы функции ν = ν(ξ) увеличиваются, а минимумы уменьшаются. Это ведет к тому, что интегральная поляризация возрастает (Рис.6). Отсюда следует, что при данной ориентации магнитного поля **H** имеет место усиление ФМЭ эффекта в исследуемой плёнке, причем степень усиления ослабевает с возрастанием величины λ. В то же время, как видно из Рис.6, критическое поле $λ_n$ увеличивается.

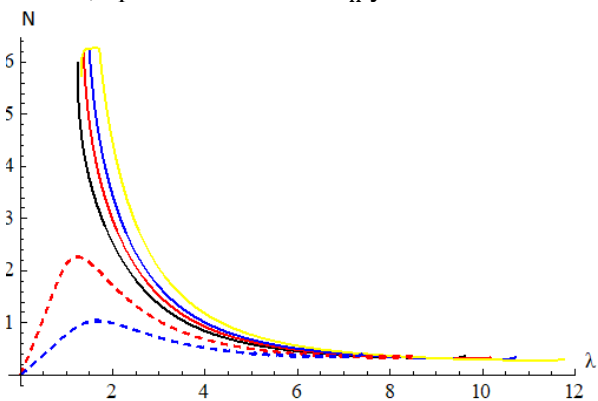

Рис.6. Графики зависимостей интегральной поляризации 0° ДГ (I) от параметра λ в магнитном поле H ∥ Oz. Здесь черная линия соответствует h = 0, красная - h = 0.2, синяя - h = 0.4, желтая h = 0.8, красная штриховая - h = -0.4, синяя штриховая - h = -0.8.

В ситуации, когда магнитное поле **H** и ось Oz антипараллельны, конфигурация профиля 0° ДГ (I) также не меняется, но изменения в топологии, которые всё – таки имеют быть, носят характер, противоположный тому, что было отмечено для случая поля **H**, направленного вдоль оси Oz. В частности, с возрастанием h интегральная поляризация N уменьшается, а также резко уменьшается поле зарождения 0° ДГ (I) (при h=0,1, $λ_n = 0$). Эта означает, что при действии магнитного поля **H**, совпадающей с легкой осью одноосной анизотропии, но направленной противоположно электрическому полю, зарождение 0° ДГ (I) возможно при произвольном λ> 0. Таким образом, действие магнитного поля с данным направлением приводит к исчезновению порога зарождения 0°ДГ в неоднородном электрическом поле, а также к ослаблению ФМЭ эффекта в рассматриваемом образце.

В случае, когда **H**∥Oy структура 0° ДГ (I) существенно трансформируется: характер распределения вектора намагниченности становится ассиметричным (Рис.7).

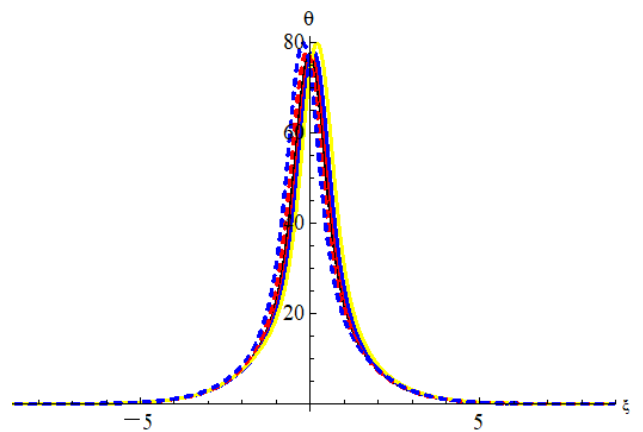

a)

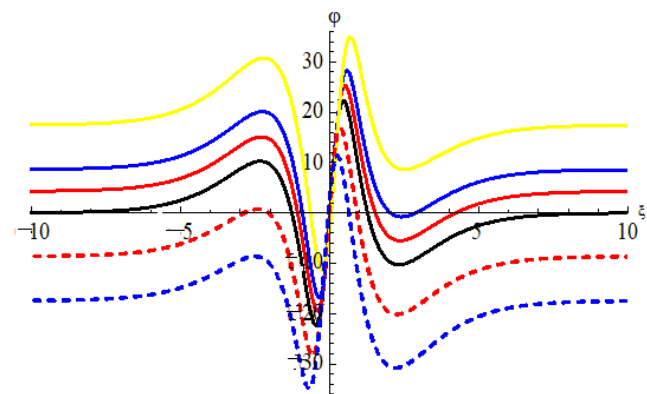

b)

Рис.7. Графики зависимостей углов θ (a) и φ (b) $0^0$ ДГ (I) от координаты ξ в магнитном поле H ∥ Oz, λ = 4. Черная линия соответствует h = 0, красная - h = 0.2, синяя - h = 0.4, желтая h = 0.8, красная штриховая - h = -0.4, синяя штриховая - h = -0.8.

В данном случае вектор **M** также отклоняется в сторону поля, но в рассматриваемой геометрии это приводит к тому, что в доменах θ(∞) = 0, φ(∞) = φ(−∞) = $φ_0$ ≠ 0. При возрастании h магнитные моменты стремятся ориентироваться вдоль поля **H** так, что величина θ(∞) остается неизменным, а φ(∞) и φ(−∞) возрастают. В то же время $θ_m$ увеличивается (хотя незначительно) и значения ξ = ξ₁, при которых θ(ξ₁) = θ_m смещаются в сторону возрастающих значений ξ. Такая же тенденция имеет место на графике зависимости φ = φ(ξ), где максимумы и минимумы φ(ξ) смещаются в ту же сторону (Рис.7). Кроме того при возрастании h графики смещаются вверх, т.к. при этом величины φ(−∞) и φ(∞) также возрастают. В итоге конфигурация магнитных моментов 0° ДГ изменяется и создает ассиметричную картину распределения намагниченности. Соответственно, ассиметричным становится и распределение поляризации по толщине ДГ. При этом с увеличением h максимумы и минимумы профиля поляризации ν = ν(ξ) по абсолютной величине увеличиваются и смещаются относительно оси ординаты в сторону возрастающих значений ξ. Следует отметить также, что критическое поле зарождения $\lambda_n$ с увеличением поля h также увеличивается (Рис.8), а начальное значение интегральной поляризации $N_n$ (её величина в момент зарождения 0° ДГ (I) при этом понижается. Однако её величина при фиксированном значении поля λ при возрастании h немного повышается (Рис.8). Таким образом при данной ориентации поля **H** ФМЭ эффект усиливается, но в незначительной степени.

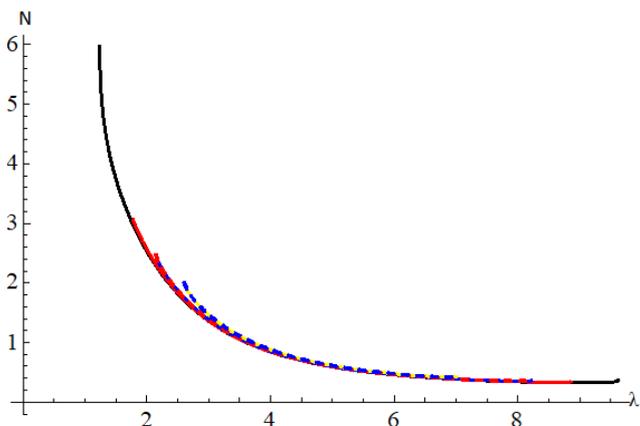

Рис.8. Графики зависимостей интегральной поляризации 0° ДГ (I) от параметра λ в магнитном поле H ∥ Oy. Черная линия соответствует h = 0, красная - h = 0.2, синяя - h = 0.4, желтая h = 0.8, красная штриховая - h = -0.4, синяя штриховая - h = -0.8.

В случае действия магнитного поля в обратном направлении, графики зависимости θ = θ(ξ) и φ = φ(ξ) полностью повторяют ход соответствующих кривых для случая положительных значений поля с той лишь разницей, что они смещаются относительно оси ординат в сторону отрицательных значений ξ. Поэтому и выводы о характере проявления ФМЭ эффекта для данной ориентации будут такими, что и в предыдущем случае.

## 5. 0° ДГ неелевского типа

Рассмотрим ещё одно решение уравнений (5), которое было найдено в работе [19] и соответствует 0°ДГ с неелевским законом изменения намагниченности (**M**: θ = 0, φ = φ(ξ). Данный тип магнитных неоднородностей весьма необычен и имеет ряд отличительных особенностей. В частности, условием его возникновения является наличие неоднородного электрического поля, в однородном поле такое решение отсутствует. Кроме того, функция φ = φ(ξ) представляет собой ограниченную нечетную функцию с двумя экстремумами, равными по величине и противоположными по знаку (Рис.9).

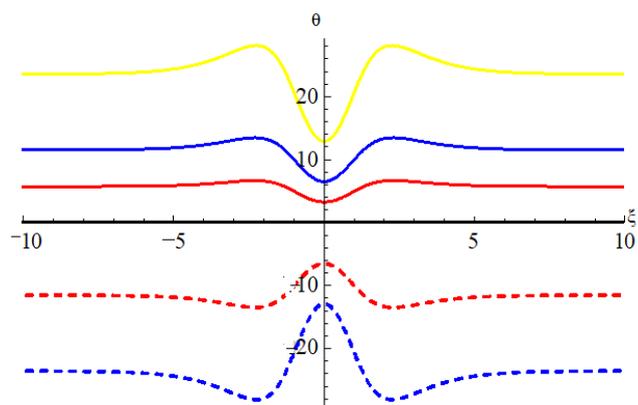

a)

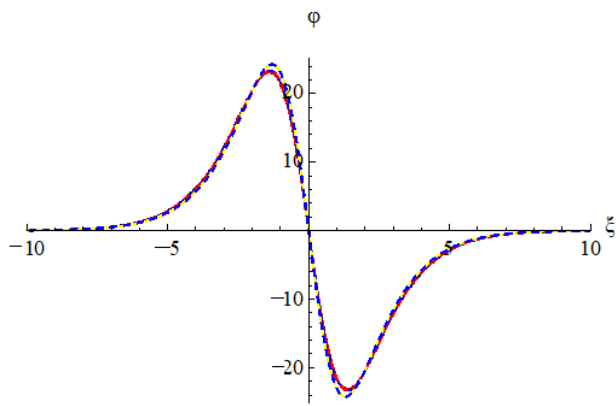

b)

Рис. 9. Графики зависимостей углов θ и φ 0⁰ ДГ (II) от координаты в магнитном поле H ∥ Ox. Здесь черная линия соответствует h = 0, красная - h = 0.2, синяя - h = 0.4, желтая h = 0.8, красная штриховая - h = -0.4, синяя штриховая - h = -0.8.

Величина φ в малых полях прямо пропорциональных λ, причем в этом случае основной вклад в зависимость φ от ξ оказывает часть неоднородного магнитоэлектрического взаимодействия, содержащая div**m** [21]. При больших значениях λ имеет место отклонение от линейного закона и основной вклад в структуру 0°ДГ (II) вносит другая часть неоднородного магнитоэлектрического взаимодействия, которая содержит rot**m**. Поляризация, возникающая в окрестности такой 0°ДГ, является четной функцией от $\xi$, не повторяющей профиль зависимости $\varphi = \varphi(\xi)$. Она имеет один максимум (в центре стенки) и два симметрично расположенных относительно центра ($\xi = 0$) минимума ($\nu_m < 0$). Отсюда следует, что распределение зарядов в 0° ДГ (II) представляет собой тройной электрический слой. Его результирующее поле, как показывают расчёты, равно нулю, т.е. в рассматриваемой ситуации имеет место полная экранировка полей, создаваемых отдельными слоями. Таким образом, интегральная поляризация N данной неоднородности равна нулю и этот результат не зависит ни от величины поля $\lambda$, ни от размера области её действия l. Таким образом, связанные заряды под действием неоднородного электрического поля индуцируется в окрестности 0°ДГ, но при этом данная неоднородность не притягивается и не отталкивается от источника электрического поля. Это означает, что ФМЭ эффект, как таковой, в данном случае никак не проявится, хотя и имеет место возникновение зарядов.

При действии магнитного поля с **H**∥O*x* на образец структура 0° ДГ (II) трансформируется следующим образом: магнитные моменты, поворачиваясь в сторону поля, способствуют выходу намагниченности из плоскости ДГ и в итоге $\theta \neq 0$. В силу того, что обменное взаимодействие на разных участках ДГ проявляется по-разному (в центре ДГ величина $\varphi'$ принимает максимальное значение, а в экстремальных точках, в которых $\varphi' = 0$, имеет минимальное значение), то появляется зависимость $\theta$ от $\xi$ (Рис. 9, a). При этом в центре ДГ ($\xi = 0$) $\theta$ принимает минимальное значение, а в экстремальных точках – максимальное. Следует отметить, что выражение для обменного взаимодействия (первое слагаемое в подинтегральном выражении (1)) содержит члены, содержащие как $(\theta')^2$, так и $(\varphi')^2$. В данном случае, зависимость $\varphi = \varphi(\xi)$ (Рис. 9, b) изменяется незначительно – лишь на немного ($< 10\%$) возрастают абсолютные значения экстремумов функции $\varphi = \varphi(\xi)$. В то же время функция $\theta = \theta(\xi)$, которая имеет минимум (при $\xi = 0$) и два симметрично расположенных максимума (относительно $\xi = 0$), с возрастанием h существенно изменяется. В частности, возрастает величина экстремумов, а разность между значениями максимума и минимума функции $\theta(\xi)$ также повышается. Кроме того увеличивается и расстояние между точками расположения максимумов. Дифференциальная поляризация $\nu = \nu(\xi)$ представляет собой также четную функцию, имеющую один максимум (в центре ДГ) и два минимума. С увеличением h величины экстремумов незначительно возрастают, а расстояние между максимумами убывает. Это в свою очередь ведёт к появлению ненулевой величины интегральной поляризации, возрастающей с повышением h, а также и $\lambda$ (Рис.10).

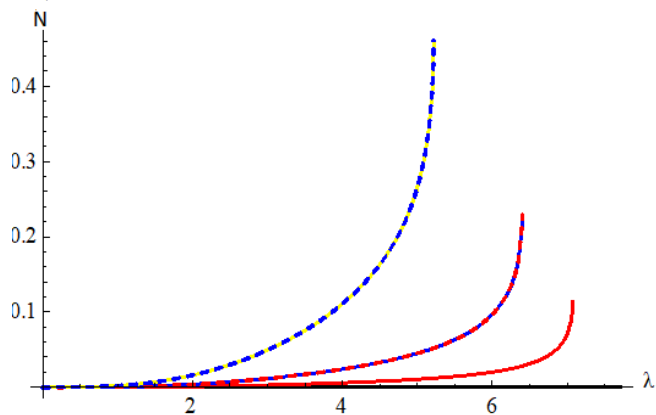

Рис.10. Графики зависимостей интегральной поляризации 0° ДГ (II) от параметра $\lambda$ в магнитном поле H ∥ O*x* . Черная линия соответствует h = 0, красная - h = 0.2, синяя - h = 0.4, желтая h = 0.8, красная штриховая - h = -0.4, синяя штриховая - h = -0.8.

Зависимость интегральной поляризации от $\lambda$ имеет вид резко возрастающей функции, стремящейся к вертикальной асимптоте $\lambda = \lambda_0$, причём чем больше поле h, тем меньше значение $\lambda_0$. Таким образом, в рассматриваемом магнетике при **H**∥O*x* возникает ФМЭ эффект, который затем (при возрастании h) может значительно усилен.

Если магнитное поле **H** направлено антипараллельно оси O*x*, то его действие приводит к наклону магнитных моментов в противоположную сторону на угол $\theta_0 = \theta(\infty) = \theta(-\infty) < 0$. За счет обменного взаимодействия появится зависимость $\theta = \theta(\xi)$ (как и в случае **H**∥O*x*), которая в итоге приводит к возникновению ненулевой интегральной поляризации N=N($\lambda$). Последняя будет иметь такой же вид, как и в предыдущем случае, т. к. плоскость O*yz* является (как элемент симметрии) плоскостью отражения рассматриваемой магнитной системы до приложения магнитного поля на образец. Соответственно, возникающий ФМЭ эффект будет иметь такие же качественные и количественные изменения топологических и физических характеристик 0°ДГ (II).

В случае **H**∥O*y* структура 0°ДГ (II) также изменяется, что связано с поворотом намагниченности в доменах в сторону направления поля на угол $\varphi_0 = \varphi(\infty) \neq 0$. Соответственно, графики функции $\varphi = \varphi(\xi)$ (Рис.11,a) с возрастанием h как бы сдвигаются вверх с одновременным увеличением экстремальных значений φ(ξ) по модулю, т.е. графики зависимости приобретают асимметричный вид. В тоже время магнитные моменты не выходят из плоскости Oyz, следовательно, θ = 0. Дифференциальная поляризация также имеет асимметричный вид, что имеет аналогию с распределением зарядов в 0°ДГ (I). В данном случае минимум функции $\nu = \nu(\xi)$(Рис. 11, b), расположенный слева от ξ = 0, с повышением h увеличивается по абсолютной величине, а правый, наоборот, уменьшается. При этом точка максимума функции ν(ξ) смещается в сторону отрицательных значений ξ. Тем не менее интегральная поляризации остается равной нулю.

лишь абсолютные значения экстремумов функции φ = φ(ξ) (Рис. 12, a). Аналогичные изменения наблюдаются и в профиле распределения индуцированных зарядов в окрестности 0°ДГ (II) (Рис.12,b).

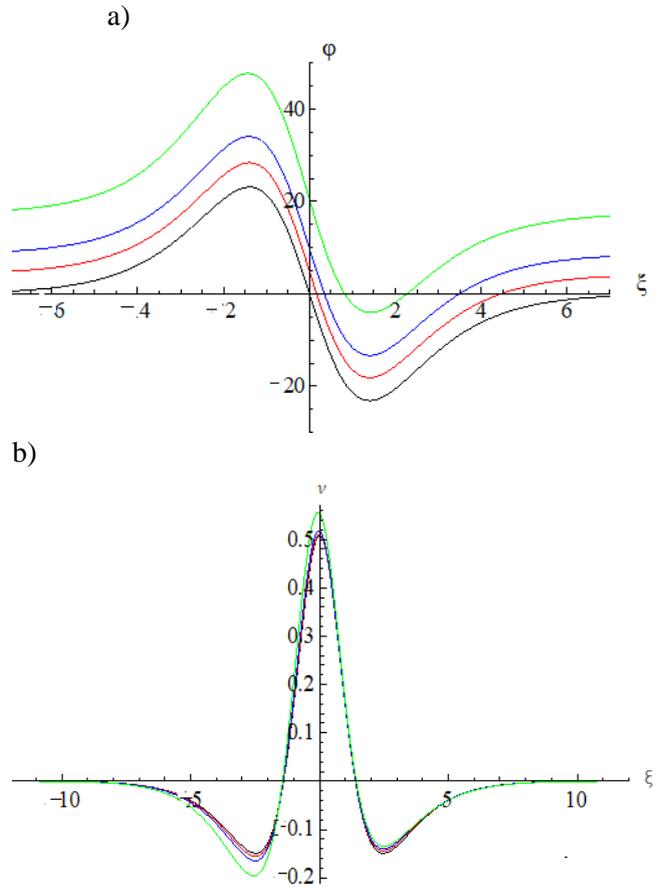

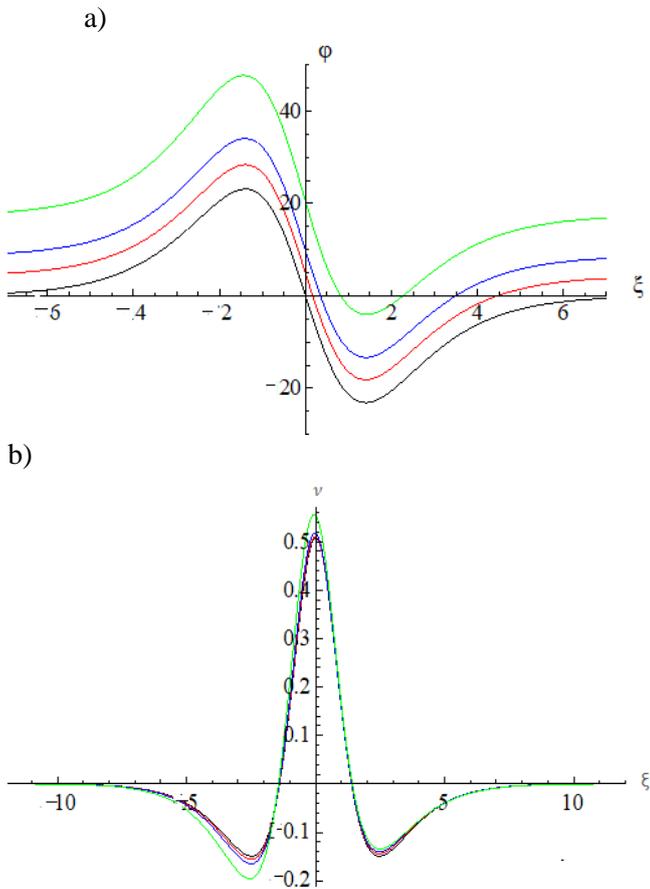

Рис. 12. Профили угла φ (a) и поляризации ν (b) в магнитном поле h ∥ Oz при λ = 4. Черная линия соответствует h = 0, красная – h =0.2, синяя – h=0.4, зеленая – h =0.8.

Рис.11. Профили угла φ (a) и поляризации ν (b) в магнитном поле h ∥ OY при λ = 4. Черная линия соответствует h = 0, красная – h =0.2, синяя – h=0.4, зеленая – h =0.8.

При действии магнитного поля, направленного противоположно оси Oy, состояние магнетика меняется также, как и в случае **H**∥Oy. В то же время характер распределения намагниченности в образце не меняется. В частности, θ = 0 и ФМЭ эффект не возникает.

Подобная ситуация имеет место и для **H**∥Oz. В данном случае симметрия распределения намагниченности не нарушается, уменьшаются

Здесь также интегральная поляризация равна нулю. Если направление магнитного поля сменить на противоположное, то ситуация с проявлением ФМЭ эффекта не изменится, т. е. будет такой же, как и в случае с **H**∥Oz.

### 5. Обсуждение результатов

Таким образом из приведенных результатов следует, что 0º ДГ обоих типов, возникающие в изучаемых магнетиках при весьма специфических условиях, существенно трансформируются при действии магнитного поля. Характер этих изменений сказывается на характеристиках 0º ДГ и определяется в основном направлением **H**. В данном случае такими параметрами 0º ДГ с точки зрения их возможного использования в технических устройствах являются интегральная поляризация N и критическое поле их зарождения $\lambda_n$. Согласно расчетам в случае 0º ДГ (I) наиболее значительное возрастание интегральной поляризации можно достичь при действии магнитного поля вдоль оси Oz, при следующих значениях параметров: $\lambda = 2$ и h=0.8 (Q=1,l=5). При этом величина N возрастает почти в 2 раза. Аналогичное усиление можно добиться и при H∥Ox, но при $\lambda = 7$, h=-0.8, т.е. при очень больших величинах электрических полей. В тоже время при данной ориентации магнитного поля возможно понижение порогового поля зарождения 0º ДГ (I) до 0 ($\lambda_n = 0$). Подобная ситуация может иметь место и для H∥Oz (h=0.1). Приведенные примеры ориентации магнитного поля, когда $\lambda_n = 0$, создают предпосылки для возможного экспериментальное наблюдения данного типа 0º ДГ.

В случае 0º ДГ (II) ситуация осложняется тем, что распределение намагниченности лежит в плоскости Oyz, которая в свою очередь является (как элемент симметрии) плоскостью отражения магнитной системы. Соответственно, действие магнитного поля, направленного вдоль оси Oy и вдоль Oz не может привести к выходу намагниченности из плоскости Oyz. Отсюда вытекает, что θ=0. При этом индуцирование зарядов происходит таким образом, что их распределение в области 0º ДГ (II) является функцией четной, а интегральная поляризация остается нулевой.

В случае H∥Ox симметрия магнитной системы нарушается, в результате появляется отклонение намагниченности от плоскости yOz, т.е. θ=θ(y). Это, в свою очередь приводит к возникновению ненулевой интегральной поляризации, которая с возрастанием λ увеличивается и при стремлении $\lambda \to \lambda_0$ резко возрастает.

Таким образом исследования показывают, что воздействием магнитного поля на магнитоодноосную пленку с неоднородным магнитоэлетрическим взаимодействием можно регулировать многими свойствами пленки, в частности, влиять на степень проявления ФМЭ эффекта в них. Более того, можно также уменьшить критическое поле зарождения 0 ДГ (I) до нуля ($\lambda_n = 0$). Последнее означает, что таким образом можно наблюдать в рассматриваемых пленках зарождение 0º ДГ. Однако, необходимо отметить, что обнаружение одномерных микромагнитных структур, которыми являются 0º ДГ выше указанных типов, в экспериментальных условиях крайне сложно осуществить, хотя бы потому, что трудно создать одномерную узкую полосу воздействия электрического поля на пленку. В реальности область воздействия поля на нее имеет 2D-геометрию (в виде пятна с круговой, эллиптической и другими формами) и, следовательно, эти неоднородности точно должны быть двумерными образованиями. В частности, если электрическое поле создается заряженной иглой [4], то наблюдаемые магнитные структуры будут представлять вихреподобные неоднородности (цилиндрические магнитные домены, магнитные скирмионы, магнитные вихри и т.д. [23]). Можно отметить, что такие структуры, в частности заряженный цилиндрический магнитный домен наблюдался в работе [23]. Для того, чтобы сказанное было более убедительным свидетельством наблюдения 0º ДГ в такой форме, необходимо провести соответствующие расчеты в двумерной модели. Однако, такие расчеты выходят за рамки нашего исследования, но в дальнейшем постараемся восполнить этот пробел.